# Exceptional surface-enhanced Rotation Sensing with Robustness in a WGM microresonator

Wenxiu Li[1], Yang Zhou[2], Peng Han[1], Xiaoyang Chang[1], Shuo Jiang[1], Hao Zhang[2*], Anping Huang[1] and Zhisong Xiao[1,3]

[1] *Key Laboratory of Micro-nano Measurement, Manipulation and Physics (Ministry of Education), School of Physics, Beihang University, Beijing 100191, China*
[2]*Research Institute of Frontier Science, Beihang University, Beijing, 100191, China*
[3]*Beijing Academy of Quantum Information Sciences, Beijing 100193, China*

Exceptional points (EPs) are special singularities of non-Hermitian Hamiltonians. At an EP, two or more eigenvalues and the corresponding eigenstates coalesce. Recently, EP-based optical gyroscope near an EP was extensively investigated to improve the response to rotation. However, the highly resultant dimensionality in the eigenstate space causing from the EP-based system is sensitivity to more external perturbations, that is, the great response at EPs requires strict implementation conditions. To solve this problem, a new non-Hermitian structure was proposed that realizes an exceptional surface (EP surface) constructed of numerous EPs embedded in a high-dimensional parameter space. With respect to the isolated EPs, non-Hermitian EP surface provides additional degrees of freedom to permit the undesired perturbations (such as fabrication errors) shifts along the EP surface which corresponds to the robustness. On the other hand, the rotation will force the system away from these EPs obtaining a sensitivity enhancement of three orders of magnitude, compared to the traditional Sagnac effect rotation detecting. The EP surface system has a potential to combine robustness with enhanced sensitivity of the rotation.

## 1. Introduction

Exceptional points (EPs), arising in non-Hermitian Hamiltonians, are peculiar singularities, where two or more eigenvalues and corresponding eigenstates coalesce, and have been theoretically and experimentally demonstrated on various non-Hermitian platforms [1-6]. Optical microcavities as ideal platform to exhibit EPs were subsequently demonstrated in the non-Hermitian studies, typically for parity-time (PT) symmetry[1, 6-9], anti-parity time (APT) symmetry[10-12] and the chiral behaviors[13, 14]. Rich physical phenomena has been observed around this point, especially the ultra-high sensing potential[14-18]. Difference from the conventional diabolic points (DPs) sensing that response is linearly proportional to the perturbation strength ε, the system located at the Nth-order EP is drastic to external perturbations, since the response induced by the perturbation scales as $\varepsilon^{1/N}$ [17, 19].

Non-Hermitian optical gyroscope under EPs exhibits substantial ramifications on sensitivity enhancement of rotation detection. Generally, resonant optical gyroscopes are (ROG) based on the conventional Sagnac effect, namely the resonant difference-frequency of the counterpropagating travelling waves (CW and CCW waves) in a resonator is linearly changed with the rotational velocity $\Omega$[20-22]. The PT-symmetric ring laser gyroscope (RLG) was first investigated, in theory when the gyroscope is tailored at EPs the scale factor of the rotation sensitivity is eight orders of magnitude enhancement than conventional Sagnac effect[23]. However, the enhancement of PT-symmetry gyroscope is reflected in the complex eigenfrequency splitting with the output spectrum broadening, which is not easier to readout. Intriguingly, APT-symmetric gyroscope structure was designed and exhibited an entire real frequency splitting at the transmission spectrum with the rotation[24]. The experiment version of non-Hermitian optical gyroscope has been successfully built through making the Brillouin laser gyroscope near the EP, they investigated the predicted EPs-enhanced Sagnac effect and observed a four-fold enhancement in the Sagnac scale factor[25]. In the same period, Khajavikhan judiciously modified the commercial Helium–Neon RLG (ring laser gyroscope) to make it operate at EP. At this point, the RLG responsing to the rotational speed has a square-root enhancement by up to a 20 times scale factor compared to the conventional Sagnac effect[26]. Despite EPs-enhanced rotation sensors tremendous advantage in improvement of sensitivity, all the theorical structures and practical experiment based on EPs sensing considered in the above studies have to suffer from the fabrication errors or experimental uncertainties.

Generally, the strong response locating at EPs requires strict implementation conditions to control multiple parameters simultaneously. However, the more resultant dimensionality in the eigenstate space, the system is sensitive to more external perturbations even including the undesirable physical quantities to be measured. Hence, reducing resultant dimensionality in the eigenstate space corresponds to the better robustness, therefore, a solution is to construct numerous EPs embedded in a high-dimensional parameter space.

The works on building of exceptional hypersurface (EP surface) are based on the Chiral EP (CEP) [27, 28]. CEP arises from the unbalanced contribution of clockwise (CW) and counter-clockwise (CCW) travelling modes in a microcavity, where the CW and CCW travelling wave modes coalesces into only one mode. Numerous EPs are embedded in the EP surfaces, through tailoring the system's response such that fabrication errors and experimental uncertainties make the system's working point always shift along the EP surface, then the robustness can be achieved [28]. The desirable signal perturbation that can induce the bidirectional coupling between the two travelling modes, resulting in the modes splitting of sufficiently small perturbation at the EP is larger than that at the DP. However, this kind of CEP is not appropriate for detecting the perturbation induced by a rotation due to absence of the coupling between the two modes, the modes splitting at the EP is not enhanced [16].

In this paper, we proposed a non-Hermitian structure that realizes an EP surface-enhanced rotation sensing. Numerous EPs are embedded in high-dimensional parameter space, provides additional degrees of freedom compared to isolated EPs, which can be exploited to combine robustness with enhanced sensitivity of the rotation.

# A non-Hermitian EP surface system for enhancing rotation detection

In a whispering-gallery mode (WGM) microresonator, the asymmetric backscattering of counter-propagating optical modes can be related to the chiral EPs. At an EP, one of the backscattering-induced coupling coefficients is zero, and the other is nonzero, which means that there is a fully asymmetric backscattering of the counter-propagating optical modes. Therefore, to realize EP surface-enhanced the rotation detection, we consider a new non-Hermitian photonics system as shown in Fig1. (a), where a WGM microresonator is coupled to a fiber-taper with coupling strength $\kappa$. The microresonator with the intrinsic loss rate $\gamma_0$ supports WGM modes with the resonant frequency $\omega_0$. A phase-shifted fiber Bragg grating (FBG) was introduced on end of the fiber, with a reflectivity $r_R$ as the function of the resonance frequency $\omega$ of the WGM resonator. A partial reflector is on the other side of the fiber.

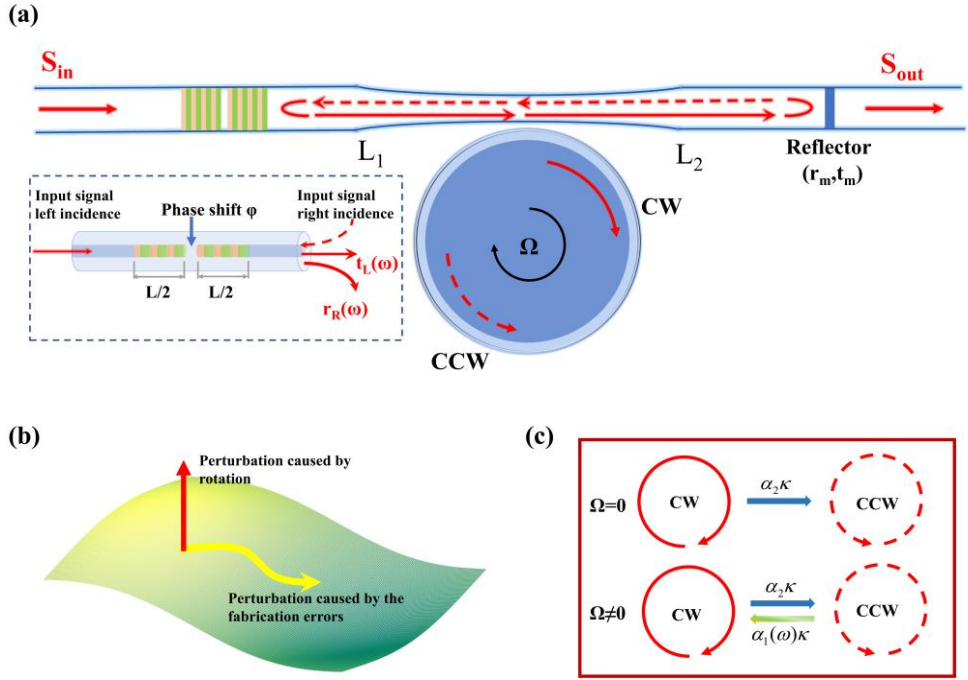

Fig.1 (a) Schematic of a non-Hermitian photonics structure that consists of a WGM microresonator coupled to a fiber-taper that has a reflector on one side and a phase-shifted FBG on the other side. Schematic of phase-shift FBG is showed in the blue box. Here, the Input signal left incidence: $S_{in}$; the Input signal right incidence: CCW mode. The reflectivity $r_R(\omega)$ and the transmittivity $t_L(\omega)$ of the phase-shifted FBG are the function of the resonance frequency $\omega$. (b) A schematic diagram of EP surface [28]. Comparison with the isolated EPs, the EP surface exhibits the characteristic that undesired perturbations due to fabrication imperfections will shift along the surface, leaving the system at an EP. (c) The optical modes coupling situation in the absence or presence of the rotation. CW and CCW optical modes in WGM microresonator are represented by a red dotted circle and a red solid circle, respectively. When the system is on rest, the $r_R$ was set to zero, the structure exhibits a fully asymmetric internal backscattering between the CW and CCW modes, which represents that the system is tailored at EP surface. In the presence of rotation, the change of the $r_R$ will induce a bidirectional coupling between the CW and CCW modes and force the system away from the EP.

Within the context of the coupled theory, the structure can be described by the effective Hamiltonian

$$H_0 = \begin{bmatrix} \omega_0 - i\gamma & \alpha_1(\omega)\kappa \\ \alpha_2\kappa & \omega_0 - i\gamma \end{bmatrix}, \quad (1)$$

where $\gamma$ is the effective loss rate of the resonator (i.e., $\gamma=\gamma_0+\kappa$ -g, g is the external gain). In addition, the unidirectional coupling strength from the CW to the CCW mode is $\alpha_1\kappa$, where $\alpha_1=r_R(\omega)\exp(2i\phi_1)$, $ir_R$ is the field reflection coefficient at the phase-shifted FBG and the propagating phase in fiber is $\phi_1=\beta L_1$ with the left arm length $L_1$, in which $\beta$ is the propagation constant of the fiber. The unidirectional coupling strength from CW mode to CCW mode is $\alpha_2\kappa$, here, $\alpha_2=r_m\exp(2i\phi_2)$, $ir_m$ is the field reflectivity of the reflector, $\phi_2=\beta L_2$ is the propagating phase in fiber with the right arm length $L_2$. Note that for setting the system at EP, a fully asymmetric backscattering between the counterpropagating optical modes should be tuned. That is, one of the backscattering strengths that are described by the off-diagonal matrix element $\alpha_1\kappa$ or $\alpha_2\kappa$ should be zero [16, 28]. In our structure, the matrix element $\alpha_1\kappa$ is tuned to zero when the system is on rest, now, the structure realizes an entire asymmetric backscattering of the counter-propagating optical modes, indicating that the backscattering from CCW to CW mode is zero and the backscattering from CW to CCW mode is non-zero. With the coupled mode theory, the effective Hamiltonian of the EP surface-based photonics structure is

$$H_{ES} = \begin{bmatrix} \omega_0 - i\gamma & 0 \\ \alpha_2\kappa & \omega_0 - i\gamma \end{bmatrix}, \quad (2)$$

hence, the corresponding eigenvalues are $\omega_{ES1,2}=\omega_0 - i\gamma$ and eigenvectors are $\tilde{a}_{ES1,2}=(0,1)^T$. According to eigenvalues, the robustness of the EP surface can be explained as, any variations of the coupling coefficient $\alpha_2\kappa$ or the effective loss rate $\gamma$ of the resonator will still leave the system at an EP. That is, there is hypersurface spanned by all possible values of these parameters in which the system remains at an EP. The EP surface-based sensors provide unprecedented robustness that cannot be achieved in standard non-Hermitian sensors depending on isolated EP.

When the structure experiences an angular rotation rate of $\Omega$ (in the CW direction), CW and CCW mode experiences opposite Sagnac frequency shift $\Delta\omega_{sagnac}=4\pi R\Omega/(n_g\lambda)$ induced by the Sagnac effect, where R is the resonator radius, $n_g$ is the group index and λ is the wavelength of the input laser. The resonance frequency of the CW mode and CCW mode change to ω+Δω$_{sagnac}$ and ω-Δω$_{sagnac}$, respectively. It is worth emphasizing that if the reflectivity of $r_R$ that does not change with the rotation, when the system rotates the Eq.(2) changes to this formula

$$H_{ES,rot} = \begin{bmatrix} \omega_0 + \Delta\omega_{sagnac} - i\gamma & 0 \\ \alpha_2\kappa & \omega_0 - \Delta\omega_{sagnac} - i\gamma \end{bmatrix}$$, however, the corresponding

eigenvalues is $\omega_{ES1,2} = \omega_0 - i\gamma \pm \Delta\omega_{sagnac}$ and the corresponding mode splitting with rotation is $\Delta\omega_{ES} = 2\Delta\omega_{sagnac}$ that distinctly shows linear relationship with rotation rate rather than the topological properties of the square root in the vicinity of EP [25, 26, 28]. Clearly, the rotation induced frequency mismatch between CW and the CCW modes will not push the system away from the EPs but shifts along the EP surface to two different EPs (i.e., ω$_{ES1,2}$), which is the reason that the chiral EP is not enhance the mode splitting induced by the rotation. Only perturbations that introduce additional coupling (i.e., changing the off-diagonal matrix element α$_1$κ or α$_2$κ) between the two modes can drastically affect the system performance. Interestingly, in our structure, the rotation induces the variation of the reflectivity $r_R$ and generates the bidirectional coupling between the CW and CCW mode, pushing the system away from the EP at last.

## 3. The reflectivity of the phase-shifted fiber Bragg grating (FBG) with the rotation

In Fig.1(a), we consider a fiber Bragg grating with refractive index distribution n(z), in which a discrete phase φ is introduced in the middle of the periodic FBG structure. The case φ= 0 corresponds to a traditional FBG without any phase shift. The grating is divided into two cells with the refractive index distribution $n(z) = n_{eff} + \Delta n_{ac}\cos(2\pi z/\Lambda + \Phi(z))$, where n$_{eff}$ is the effective refractive of the fiber, Δn$_{ac}$ is the refractive modulation strength of the grating, Λ is the nominal period, z is the position along the fiber and Φ(z) is the additional phase modulation. For a single-mode FBG, the coupling coefficient κ$_0$ of the grating is $\kappa_0 = \frac{\pi\Delta n_{ac}}{\lambda}$, The propagation constant detuning δ is the detuning from the Bragg wavelength is $\delta = 2\pi n_{neff}(\frac{1}{\lambda} - \frac{1}{\lambda_B})$, $\lambda_B = 2n_{neff}\Lambda$ is the Bragg wavelength that determines centered of the transmission resonance. The phase-shifted FBG can be described using the transfer matrix method[29-32]. In Fig.1(a), the phase-shifted FBG is utilized to reflect the CCW mode to the CW mode. Thus, the reflectivity and the transmissivity of the phase-shifted FBG responding to the Sagnac effect $\Delta\omega_{sagnac}$ can be obtained as

$$r_R = \frac{\kappa_0[\frac{n_{eff}}{c}(\Delta\omega_B - \Delta\omega_{sagnac})(e^{i\varphi}-1)r_1 - i\sigma r_1(1+e^{i\varphi})]}{-\kappa_0^2(1+r_1^2 e^{i\varphi}) + (\frac{n_{eff}}{c}(\Delta\omega_B - \Delta\omega_{sagnac}))^2(1+r_1^2) - 2i\frac{n_{eff}}{c}(\Delta\omega_B - \Delta\omega_{sagnac})\sigma r_1},$$

$$t_L = \frac{e^{i\varphi/2}\sigma^2 \text{sech}^2(\sigma l)}{-\kappa_0^2(1+r_1^2 e^{i\varphi}) + (\frac{n_{eff}}{c}(\Delta\omega_B - \Delta\omega_{sagnac}))^2(1+r_1^2) - 2i\frac{n_{eff}}{c}(\Delta\omega_B - \Delta\omega_{sagnac})\sigma r_1},$$

(3)

here, $\Delta\omega_B = \omega - \omega_B$, $\omega_B = 2\pi c/\lambda_B$, c is the optical velocity in vacuum, $\sigma = \sqrt{\kappa_0^2 - \delta^2}$, $r_1 = \tanh^2(\kappa_0 l)$, $l = L/2$. The case $\varphi = \pi$ corresponds to a π-phase shifted FBG, the hallmark of it is the narrow transmission resonance in the middle of the totally reflecting bandgap [33]. The incorporation of π phase-shifts gives rise to a narrow band transmission window inside the stopband of an FBG. Fig.2(a) and Fig.2(b) display the reflectivity $r_R$ and transmissivity $t_L$ of the π-shifted FBG with the variation of the $\Delta\omega_B$, respectively. Evidently, the reflectivity of the π-shifted FBG is narrower, in this case, the resonance frequency variation induced by the Sagnac effect will bring about a significant reflectivity change of the π-shifted FBG. To make the system locate at EP, the reflectivity $r_R$ needs close to zero in a certain resonance frequency (i.e., $\Delta\omega_B = 0$). Fig.2(c) is depicted at the case in $\omega_0 = \omega_B$, in the absence of the rotation, i.e., $\Omega = 0$, the reflectivity $r_R$ is close to zero that represents the systemin the vicinity of EP in resonance condition. With the presence of rotation, the variation of resonance frequency induced by rotation causes the variation of coupling coefficient $\alpha_1\kappa$, i.e., the bidirectional coupling between the CW and CCW modes generates, pushing the system away from the EP. Fig.2(d) shows that the slope of the reflectivity $r_R$ of the π-shifted FBG and the conventional FBG with the variation of $\Omega$.

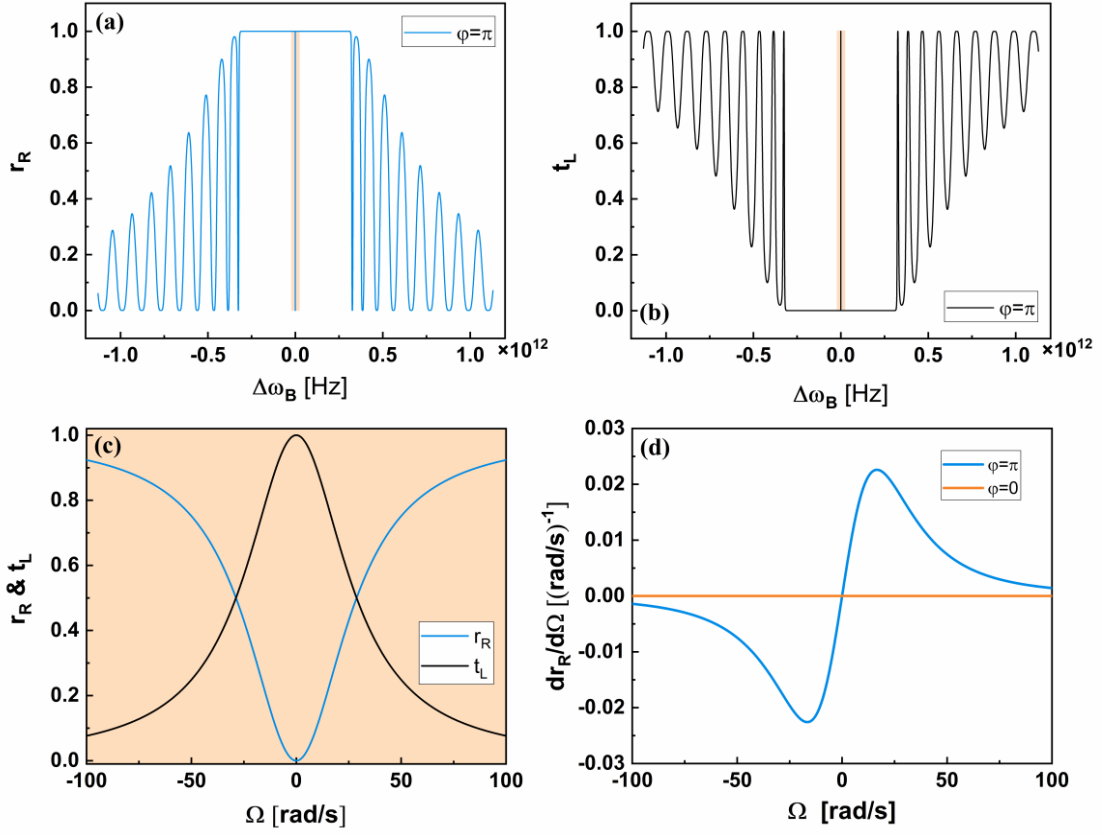

Fig.2 (a) the reflection and (b) transmission spectra of a π-shifted FBG with an effective index $n_{eff}$=1.44, a uniform index modulation $\Delta n_{ac}$ =5 × $10^{-4}$, a nominal period Λ = 538.19 nm and length $l$= 8mm. (c) is the reflection and transmission spectra of π-shifted FBG as a function of the rotation rate when the resonance wavelength of the resonator is equal to the Bragg wavelength. (d) The slope of the reflectivity $r_R$ of the π-shifted FBG and the conventional FBG with the variation of Ω.

## 4. The eigenmodes splitting of Exceptional surface System

When the system rotation with Ω, the corresponding reflectivity $r_R$ of π-shifted FBG changes, causing the bidirectional coupling between the CW and CCW modes. The Hamiltonian in rotation is given by

$$H_{0,rot} = \begin{bmatrix} \omega_0 - i\gamma + \Delta\omega_{sagnac} & \alpha_1(\Omega)\kappa \\ \alpha_2\kappa & \omega_0 - i\gamma - \Delta\omega_{sagnac} \end{bmatrix}, \quad (4)$$

The resultant frequency splitting generates, therefore, the corresponding eigenvalues of the Hamiltonian $H_0$ in Eq. (4) can be expressed by

$$\omega_{\pm} = \omega_0 - i\gamma \pm \sqrt{\kappa^2 r_R(\Omega) r_m e^{2i(\phi_1+\phi_2)} + \Delta\omega_{sagnac}^2}, \quad (5)$$

where $\Delta\omega_{\pm} = \omega_{\pm} - \omega_0$ is the eigenfrequency detuning, $r_m\kappa$ is the effective unidirectional coupling strength from the CW to the CCW mode and the $r_R(\Omega)\kappa$ subjected to rotation is the effective unidirectional coupling strength from the CCW to the CW mode. The complex eigenfrequency splitting is

$$\Delta\omega_{sp} = \omega_+ - \omega_- = 2\sqrt{\kappa^2 r_R(\Omega) r_m e^{2i(\phi_1+\phi_2)} + \Delta\omega_{sagnac}^2}. \quad (6)$$

The total coupling between two counter-propagating modes are the coherent superposition of two coupling modes for an EP surface-based structure[34, 35]. From Fig.3, the complex eigenfrequency splitting changes periodically from minimum (maximum) to maximum (minimum) with the variations of the $\phi_1$ and $\phi_2$, as predicted by the Eq. (7). The trends of the variations in the real and imaginary frequency splitting are opposite. In our system, the propagating phases in fiber are set as $\phi_1=0$ and $\phi_2=\pi$, hence, the eigenfrequency splitting exhibits an entire real splitting $\Delta\omega_{sp}=2\sqrt{\kappa^2 r_R(\Omega) r_m + \Delta\omega_{sagnac}^2}$, that an eigenfrequency splitting scaling fitting $\sqrt{r_R}$ should take place, exhibiting the character of enhanced sensitivity near a second-order EPs. Through adjusting the propagation phases in the fiber, the complex frequency splitting close to EP is avoid, which is suited to estimate the precision of EP surface sensor.

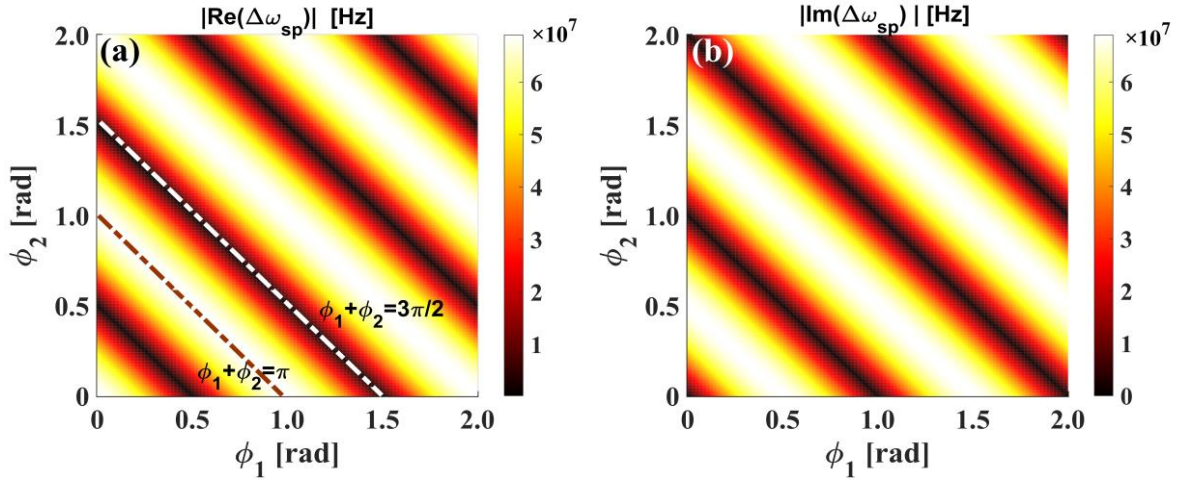

Fig. 3 Numerical simulations of the dependence of the complex eigenfrequency splitting on the phases $\phi_1$ and $\phi_2$ with rotation rate $\Omega=5$ rad/s. (a) The real part Re ($\Delta\omega_{sp}$) of the eigenfrequency splitting; (b) The imaginary part Im ($\Delta\omega_{sp}$) of the eigenfrequency splitting.

As depicted in Fig. 1(c), the proposed schematic is that a WGM resonator coupled to a fiber with a phase-shifted FBG at the one port and a reflector at the other port. Here, we consider CW and CCW modes of the WGM microresonator with the intracavity mode fields $a_{cw}$ and $a_{ccw}$, respectively. Assuming that the structure is probed via the fiber by an input signal $S_{in}$. The Heisenberg equations of motion describe the evolution of the propagating CW and CCW optical modes,

$$\dot{a}_{cw} = -(i\Delta\omega_1 + \gamma)a_{cw} - i\alpha_1\kappa a_{ccw} - it_p\sqrt{\kappa}e^{i\phi_1}s_{in},$$
$$\dot{a}_{ccw} = -(i\Delta\omega_2 + \gamma)a_{ccw} - i\alpha_2\kappa a_{cw} - it_p r_m\sqrt{\kappa}e^{i(\phi_1+2\phi_2)}s_{in}, \quad (7)$$

where $\Delta\omega_1=\Delta\omega+\Delta\omega_{sagnac}$ and $\Delta\omega_2=\Delta\omega-\Delta\omega_{sagnac}$ are the frequency detuning of the CW and CCW modes, respectively, and $\Delta\omega=\omega-\omega_0$ is the resonance frequency detuning. Based on the input-output relation, the CW and CCW output fields are described by $a_{cw}^{out}=\sqrt{\kappa}e^{i2\phi_2}a_{cw}$ and $a_{ccw}^{out}=\sqrt{\kappa}e^{i(2\phi_1+\phi_2)}a_{ccw}$, respectively. The corresponding transmission coefficient is obtained as

$$\frac{S_{out}}{S_{in}}=\frac{t_m t_{cav}(\kappa^2\alpha_1(\alpha_2-r_m r_R e^{i(\phi+\phi')})+\mathcal{R}_1\mathcal{R}_2-iA\kappa r_R e^{i\phi_1}\mathcal{R}_2+e^{i\phi'}(-e^{i\phi_1}\alpha_2\kappa^2-i\kappa r_m \mathcal{R}_1))}{\kappa^2\alpha_1\alpha_2+\mathcal{R}_1\mathcal{R}_2}, \quad (8)$$

where, $\mathcal{R}_1=\gamma+i\Delta\omega_1$, $\mathcal{R}_2=\gamma+i\Delta\omega_2$, $\phi=\phi_1+2\phi_2$, $\phi'=2\phi_1+\phi_2$. The transmission spectrum is dominated by $T=|S_{out}/S_{in}|^2$. Fig.4(a) plots a series of transmission spectrum depend on rotation rate $\Omega$. In case of non-rotation, the system exhibits a fully asymmetric backscattering at the EP, that is, the resonator supports only one travelling mode (for instance CW mode in our structure) that does not lead to frequency splitting. The resultant frequency splitting arises from rotation, with the increases of the rotation rate, the splitting of the transmission spectrum increases simultaneously, because the increase of the rotation increases the coupling coefficient from the CCW mode to CW mode. The strength of the two resonant peaks are different that can be explained by the coupling between different wave components. Due to that the backscattering-induced coupling strength ($\alpha_1\kappa$) from CW to CCW mode is not equal to that strength ($\alpha_2\kappa$) from CCW to CW mode, the two eigenmodes contain different CCW and CW components, therefore the strength of the two peaks are different. For the two eigenmodes affected by the rotation, the more they depart from each other, the less they influence each other, thus the peaks decrease as the splitting of the two eigenmodes. The frequency splitting has a nonlinear relationship with the rotation rate $\Omega$ is showed in Fig.4(b). The red and white dotted line correspond to the splitting eigenfrequency with the rotation rate $\Omega$. The transmission peaks trajectory of the system is also plotted by the contour map in Fig.4(b) that is consistent with the trajectory of the eigenfrequency splitting. In principle, the EP surface-induced sensitivity enhancement is expected to be higher for the small rotation rate. However, this is limited by the finite linewidths of resonance that can be improved further by increasing quality factors through external gain introduced.

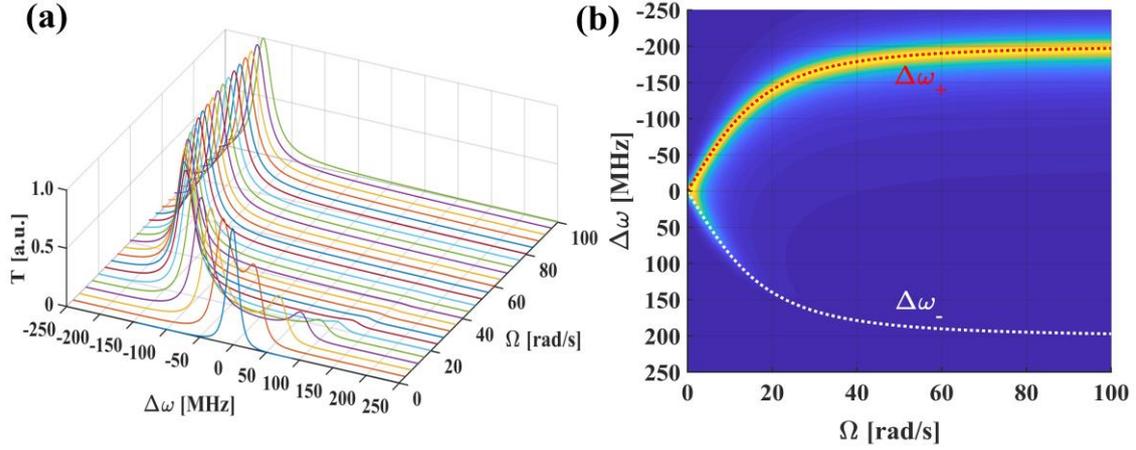

Fig.4 (a) A series of normalized transmission spectra of the EP surface-based structure as a function of rotation rate $\Omega$. (b) The trajectory of the transmission peaks as the function of the rotation rate $\Omega$ in $\phi_1=0$ and $\phi_2=\pi$. The red dotted line and the white dotted line correspond to the eigenfrequency as the function of the rotation rate $\Omega$. The parameters of the WGM resonator are set as $\kappa=2.25\times10^8$Hz, $\gamma_0=2.25\times10^8$Hz, $\gamma=1.5\times10^7$Hz, R=164μm and the phases $\phi_1=0$, $\phi_2=\pi$.

## 5. The sensitivity enhancement of the rotation detecting on EP surface

Accordingly, the EP surface-based system can measure the rotation signal $\Omega$ through monitoring the eigenfrequency splitting $\Delta\omega_{sp}$. For comparison with the conventional measurement, the scale factor is calculated as the derivative of the eigenfrequency splitting $\Delta\omega_{sp}$ with respect to the rotation rate $\Omega$,

$$S_{ES}=\frac{\partial\Delta\omega_{sp}}{\partial\Omega}\bigg|_{\Omega\to0}=\frac{1}{2}\sqrt{r_m}\kappa\frac{1}{\sqrt{r_R}}\frac{\partial r_R}{\partial\Delta\omega_{sagnac}}\frac{4\pi R}{n_g\lambda}, \qquad(9)$$

where $4\pi R/(n_g\lambda)$ is the conventional scale factor of Sagnac effect. In this equation, $\frac{1}{2}\sqrt{r_m}\kappa\frac{1}{\sqrt{r_R}}\frac{\partial r_R}{\partial\Delta\omega_{sagnac}}$ is the EP enhancement factor. This enhancement factor originates from the steep slope of the response curve operation at EP. Also, the scale factor $S_{ES}$ depends on the external coupling coefficient of the resonator $\kappa$, reflectivity $r_R$ of the π-shifted FBG and the slope of it. Hence, for realizing the larger EP enhancement factor, the frequency variation induced by the Sagnac effect will bring about a significant reflectivity change of phase-shifted FBG (i.e. $\partial r_R/\partial\Delta\omega_{sagnac}$), thus, the larger mode splitting of transmission spectrum for the WGM resonator is observed at last. The superior performance of the EP surface-based gyroscope becomes apparent at smaller rotation velocities, as depicted in Fig.5, where the sensitivity is three orders of magnitude enhancement of the conventional Sagnac effect near the EP. Clearly, the EP surface system has a better performance than conventional system, making this device valuable for measuring small rotations.

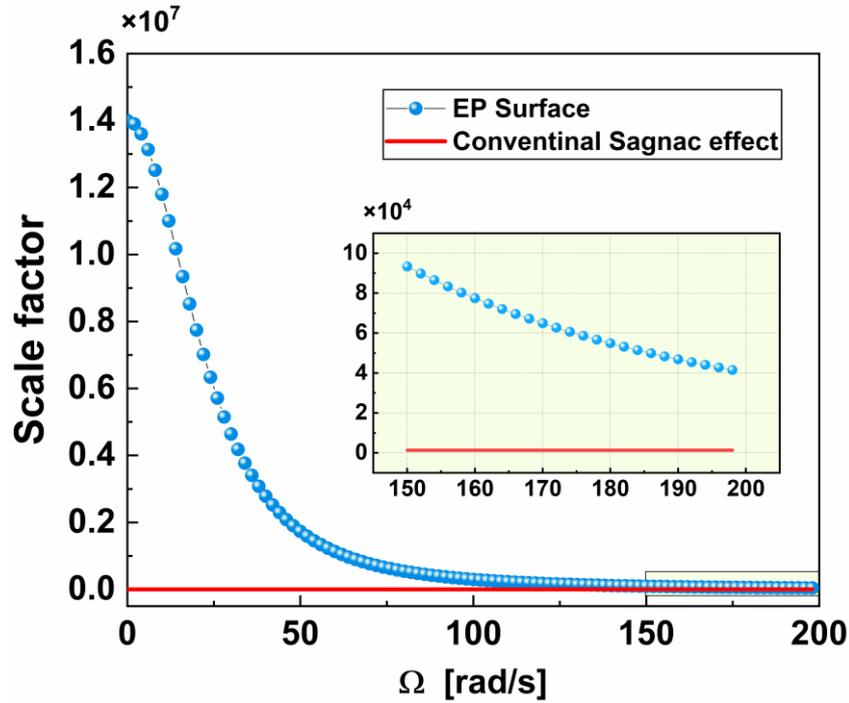

Fig.5 The Sagnac scale factor S of the EP surface system compared with the conventional Sagnac effect. The inset is amplification of the green rectangle.

The effective detection of rotation in practical is determined not only by the sensitivity (the eigenfrequency splitting) but also by the detection limit (resolvability of the two splitting eigenmodes). The rotation is detectable while frequency splitting is larger than half of the linewidths of the eigenmodes. In our system, with the increase of the external coupling coefficient, the eigenfrequency modes splitting is higher but its linewidth increases synchronously. One solution to overcoming this point is to introduce external optical gain into the resonator ($\gamma=\gamma_0+\kappa-g$). The $Q_{sp}=|\Delta\omega_{sp}/2\gamma|$ denotes the frequency splitting quality; If $Q_{sp}>1$, the the eigenfrequency splitting can be resolved easily in practical, that is, the detection limit $\Omega_{min}$ can be obtained when $Q_{sp}=1$[15, 36]. In Fig.6, the intersection of three lines and a dashed line corresponds to the detection limit in different effective loss rates, respectively. With the effective loss rate decreasing (the external gain rate increasing), the detection limit decreases.

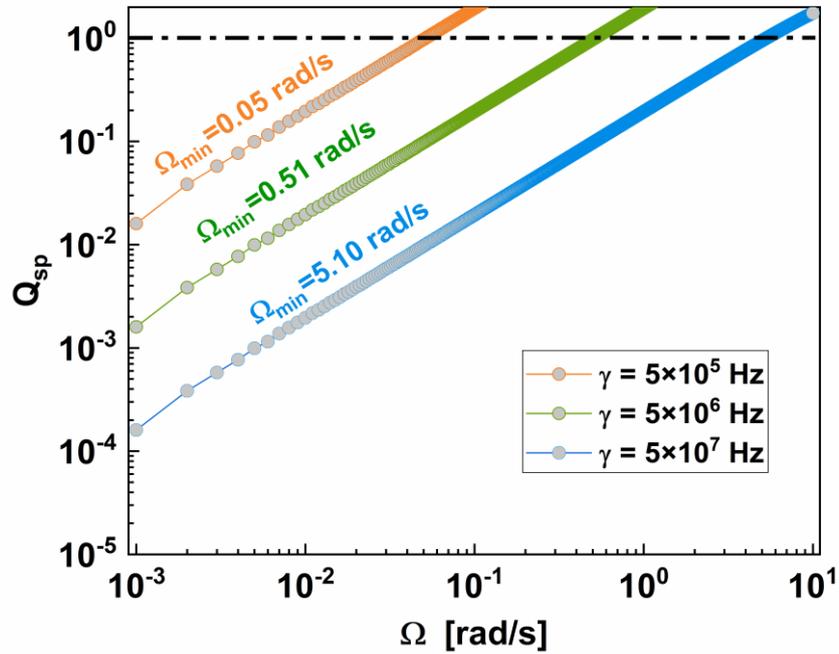

Fig.6 The frequency splitting quality $Q_{sp}$ depends on the effective loss rate. These lines show the relationship between $Q_{sp}$ and rotation rate on a logarithmic scale.

**6. Conclusions**

In conclusion, we have shown a new non-Hermitian photonics structure for an EP surface-based integrated optical gyroscope. Through adjusting the relationship between the Bragg wavelength of the FBG and the resonance frequency of the resonator, the structure can be located at EP surface. This structure operating at EP surface in place of the isolated EPs, provides additional degrees of freedom compared to isolated EPs, which can be exploited to combine a certain degree of robustness against fabrication tolerance together with the enhanced sensitivity of the rotation. The rotation will force the system away from the EP surface, causing large frequency splitting compared to the traditional Sagnac effect. The sensitivity of EP surface-based gyroscope can arise three orders of the magnitude, compared to the traditional Sagnac effect. We anticipate that the EP surface system has the potential to combine robustness with enhanced sensitivity of the rotation and will provide a series of new possibilities for sensing applications using practical non-Hermitian optical devices.

**Acknowledgement.** Financial support from the National Natural Science Foundation of China (61975005, 11804017, 11574021) and Beijing Academy of Quantum Information Sciences(Y18G28).

nanoparticle sensor," Journal of Applied Physics **128**(2020).